\documentclass[]{article}
\usepackage{color}
\usepackage{amsmath,amsfonts,amssymb}
\usepackage{graphicx}
\usepackage{jabbrv}
\usepackage{authblk}
\usepackage{url}
\usepackage{textcomp}
\usepackage{caption}
\usepackage{appendix}
\usepackage{hyperref}
\hypersetup{
   hidelinks
}
\usepackage{geometry}
\title{Ring resonator enhanced mode-hop-free wavelength tuning of an integrated extended-cavity laser}

\author[1,3]{Albert van Rees}
\author[1,2]{Youwen Fan}
\author[2]{Dimitri Geskus}
\author[2]{Edwin J. Klein}
\author[2]{Ruud M. Oldenbeuving}
\author[1,4]{Peter J.M. van der Slot}
\author[1]{Klaus-J. Boller}

\affil[1]{Laser Physics and Nonlinear Optics, Mesa$^{+}$ Institute for Nanotechnology, Department for Science and Technology, University of Twente, P.O. Box 217, 7500 AE Enschede, The Netherlands}
\affil[2]{LioniX International B.V., P.O. Box 456, 7500 AL Enschede, The Netherlands}
\affil[3]{a.vanrees@utwente.nl}
\affil[4]{p.j.m.vanderslot@utwente.nl}

\begin{document}

\maketitle

% Abstract (Do not insert blank lines, i.e. \\) 
\abstract{Extending the cavity length of diode lasers with feedback from Bragg structures and ring resonators is highly effective for obtaining ultra-narrow laser linewidths. However, cavity length extension also decreases the free-spectral range of the cavity. This reduces the wavelength range of continuous laser tuning that can be achieved with a given phase shift of an intracavity phase tuning element. We present a method that increases the range of continuous tuning to that of a short equivalent laser cavity, while maintaining the ultra-narrow linewidth of a long cavity. Using a single-frequency hybrid integrated InP-Si$_3$N$_4$ diode laser with 120~nm coverage around 1540~nm, with a maximum output of~24 mW and lowest intrinsic linewidth of 2.2~kHz, we demonstrate a six-fold increased continuous and mode-hop-free tuning range of 0.22~nm (28~GHz) as compared to the free-spectral range of the laser cavity. }

% Body
\section{Introduction}
Diode lasers with single-mode oscillation, mode-hop-free wavelength tunability and intrinsic low phase noise through a narrow Schawlow-Townes linewidth are instrumental in many applications. These include communication technology, for instance, to raise the data flow in fiber networks with advanced phase encoding~\cite{olsson_2018oe}, or as on-chip local oscillators for integrated microwave photonics~\cite{marpaung_2019np}. Other applications include retrieval of information with highest precision, sensitivity, and speed, such as with spectroscopic detection, monitoring and sensing~\cite{wieman_1991rsi, tombez_2017o, lopez_2011JLT, allen_1998mst}, for ranging with diode-driven frequency combs~\cite{stern_2018n, raja_2019natcom, suh_2016science}, or to advance time keeping with chip-integrated, portable optical clocks~\cite{newman_2019o}.

In all these cases, mode-hop-free tunability is as important as a narrow linewidth. While the latter enables precision, tunability is essential to make systematic use of the precision. This includes accessing target wavelengths, \textit{e.g.}, in the ITU grid, and controlled scanning across given wavelength intervals as in interferometry. It is also essential for electronic stabilization to absolute frequency standards~\cite{deLabachelerie_1995ol}.

At first sight, there appears to be conflicting optical requirements between obtaining ultra-narrow linewidth versus the spectral range across which mode-hop-free tuning is possible. The reason is that linewidth and tuning range both depend on the cavity length of the considered laser, however, in an opposing manner. When increasing the laser cavity length, $L_c$, this increases the cavity photon lifetime and the number of photons in the resonator, which yields an inversely quadratic reduction of the Schawlow-Townes or intrinsic linewidth, $\Delta\nu_{\textrm{ST}}~\propto~1/L_c^2$~~\cite{schawlow_1958pr,lax_1968pr, henry_1982JQE}. However, extending the cavity length also increases the longitudinal mode density, which, by itself, decreases the range of mode-hop-free tuning, $\delta \lambda_c \propto 1/L_c$, to the free-spectral range (FSR) of the laser, and can lead to multi-mode oscillation. This is typically resolved by inserting a tunable optical filter into the cavity. Narrow-linewidth, single-mode oscillation that can be mode-hop-free tuned over a large range is then obtained by tuning the lasing wavelength via moving one of the cavity mirrors and adjust the filter center wavelength accordingly~\cite{littman_1978ol,liu_1981ol,flemming_1981jqe}. 

Such a solution cannot be pursued for fully integrated lasers, as the lasing wavelength cannot be controlled by moving parts, \textit{i.e.}, by changing the geometric length of the laser cavity via displacement of a mirror. Instead, these lasers contain a phase section (PS)~\cite{kobayashi_1988jlt,coldren_2000jstqe} to change the phase of the recirculating light and, hence, its wavelength. Typically, the maximum phase tuning in integrated circuits is physically limited by the small values of material coefficients that facilitate the phase tuning, \textit{e.g.}, the thermo-optic, strain-optic or electro-optic coefficients. To nevertheless obtain a large continuous tuning range, it is highly desirable to induce a large wavelength shift, $\delta\lambda_c$, per unit phase added by the phase section, $\delta \phi_{\textrm{ps}}$. Therefore, the tuning sensitivity, defined as $F_{\lambda} \equiv \partial\lambda_c / \partial \phi_{\textrm{ps}}$, should be maximized.

\begin{figure}[tb]
	\centering
	\includegraphics[width=\textwidth]{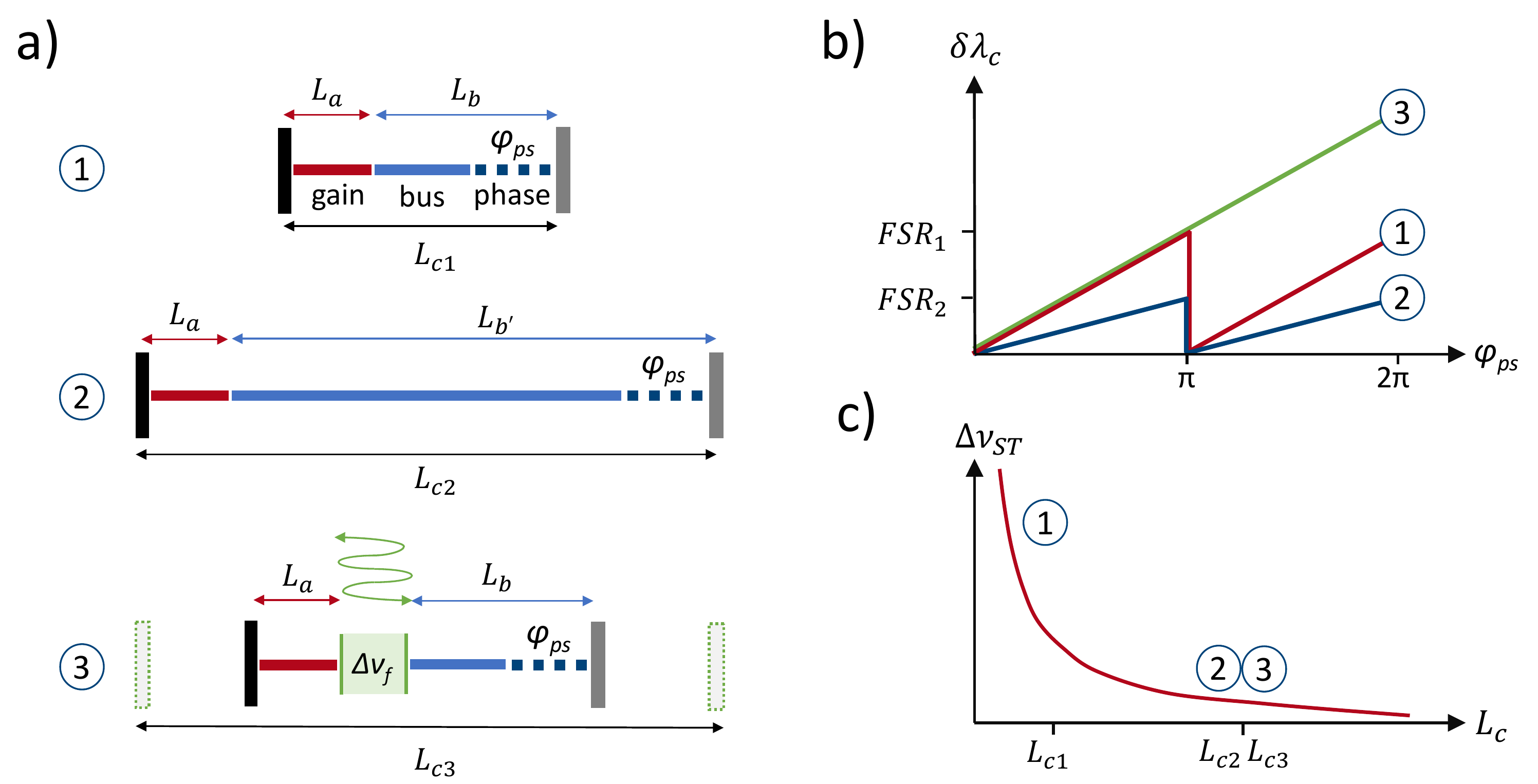}
	\caption{(a) Schematic view of three lasers having the same optical amplifier section but with different cavity configurations. Laser~(1) has a short cavity of length $L_{c1}$, including a gain element of length $L_a$ and a bus section of length $L_b$, which includes a phase tuning element $\phi_{ps}$. The bus section length is increased  for laser~(2), resulting in a cavity length $L_{c2}\gg L_{c1}$. Laser~(3), includes an optical filter with bandwidth $\Delta\nu_f$ that provides for both single-mode oscillation and significant enhancement of the cavity photon lifetime. The resulting effective cavity length $L_{c3}$ is assumed to be equal to that of laser~(2), while the bus waveguide is taken the same as that of laser~(1). (b) The tuning behavior of the three lasers. (c) The according Schawlow-Townes linewidth for the three lasers. See text for further details.}
	\label{fig:cavities}
\end{figure} 

Figure~\ref{fig:cavities} illustrates the tuning and linewidth limitations of integrated lasers and our strategy to obtain a large tuning sensitivity $F_{\lambda}$ and at the same time maintain single-frequency oscillation with a narrow intrinsic linewidth. Laser 1 in Fig.~\ref{fig:cavities}~(a) is a schematic representation of the simplest fully integrated laser which comprises a gain element of length $L_a$, and a bus section of length $L_b$ that includes a phase section. The total cavity length $L_{c1}=L_a+L_b$ can be small and therefore the laser possesses a large tuning sensitivity $F_{\lambda} \propto 1/L_{c1}$. The corresponding laser cavity free-spectral range is large, enabling continuous tuning by the phase section over a range up to $\textrm{FSR}_1$, as is schematically shown in Fig.~\ref{fig:cavities}~(b). On the other hand, such a configuration with a small cavity length results in a broad Schawlow-Townes linewidth, as is schematically shown in Fig.~\ref{fig:cavities}~(c). 

To obtain a small Schawlow-Townes linewidth, the bus waveguide can be extended to a length $L_{b^{'}}\gg L_{b}$, as is schematically shown as laser 2 in Fig.~\ref{fig:cavities}~(a). The total laser cavity length, $L_{c2}$, is now much larger than that of laser 1. Consequently, the tuning sensitivity $F_{\lambda} \propto 1/L_{c2}$ becomes smaller and the continuous tuning range is limited to the small FSR of the long laser cavity, as shown in Fig.~\ref{fig:cavities}~(b). Nevertheless, laser 2 will have a narrow Schawlow-Townes linewidth as is shown in Fig.~\ref{fig:cavities}~(c). Our demonstration experiment of such a laser with a long cavity 
will confirm both the narrow Schawlow-Townes linewidth and the small tuning sensitivity $F_{\lambda}$, when only tuning the phase section.

In contrast to this, when a filter is added to the laser cavity that significantly increases the cavity photon lifetime, the bus waveguide can be kept short, \textit{e.g.}, as short as the bus waveguide of laser~1. Such a laser is schematically shown as laser~3 in Fig.~\ref{fig:cavities}~(a). Filters that enhance the photon cavity lifetime are typically resonator based~\cite{liu_2001apl,oldenbeuving_2013lpl, kobayashi_2015jlt, fan_2016pj, stern_2017ol, Komljenovic_2017as, lin_2018pj, mak_2019oe,fan_2019arxiv,tran_2020jstqe} and rely on the light circulating multiple times within the resonator per roundtrip through the laser cavity. The multiple passes through the resonator effectively extend the cavity length, $L_{c3}$ to beyond the physical mirror spacing, as is schematically indicated by the dashed mirrors. Maximizing the contribution of the optical filter to the cavity photon lifetime ensures that the laser provides single-frequency oscillation with a narrow Schawlow-Townes linewidth. Moreover, our demonstration experiment shows that the tuning sensitivity $F_{\lambda}$ of laser~3 can be increased to that of laser~1, for the case of mode-hop-free tuning, which relies on synchronous tuning of the phase section with the resonant filter (see Fig.~\ref{fig:cavities}~(b)). This mode-hop-free tuning in combination with the increased $F_{\lambda}$ allows the continuous tuning range to extend far beyond the free-spectral range of the laser cavity. 

Specifically, for the case of a filter based on two microring resonators as used in our demonstration experiment, mode-hop-free tuning requires phase shifters in both rings to keep the resonant wavelength of the rings in synchronization with the oscillating wavelength of the laser. Under this condition, the tuning sensitivity $F_{\lambda}$ is given by (see Appendix for derivation)
\begin{equation}
F_{\lambda}\equiv\frac{\partial\lambda_{c}}{\partial\phi_{\textrm{ps}}}=\frac{1}{\pi}\frac{\lambda_{c}^{2}}{2n_{g,a}L_{a}+2n_{g,b}L_{b}}.
\label{eq:sensitivity}
\end{equation}
In Eq.~\ref{eq:sensitivity}, $\lambda_{c}$ is the lasing wavelength, $n_{g,a}$ and $L_{a}$ are the effective group index of the waveguide in the optical amplifier and its length, respectively, and $n_{g,b}$ and $L_{b}$ are the according quantities for the silicon nitride bus waveguide.

Equation~\ref{eq:sensitivity} summarizes our central finding: the tuning sensitivity $F_{\lambda}$ is independent of the optical or physical length of the microring resonators that extend the overall cavity length. Therefore, Eq.~\ref{eq:sensitivity} is also valid for any number of ring resonators. Essentially, this allows to extend the laser cavity length to huge values with many roundtrips through multiple ring resonators, to reduce the laser linewidth, without reducing the tuning sensitivity $F_{\lambda}$ when the integrated laser is mode-hop-free tuned.

\begin{figure}[t]
    \centering
    \includegraphics[width=0.57\textwidth]{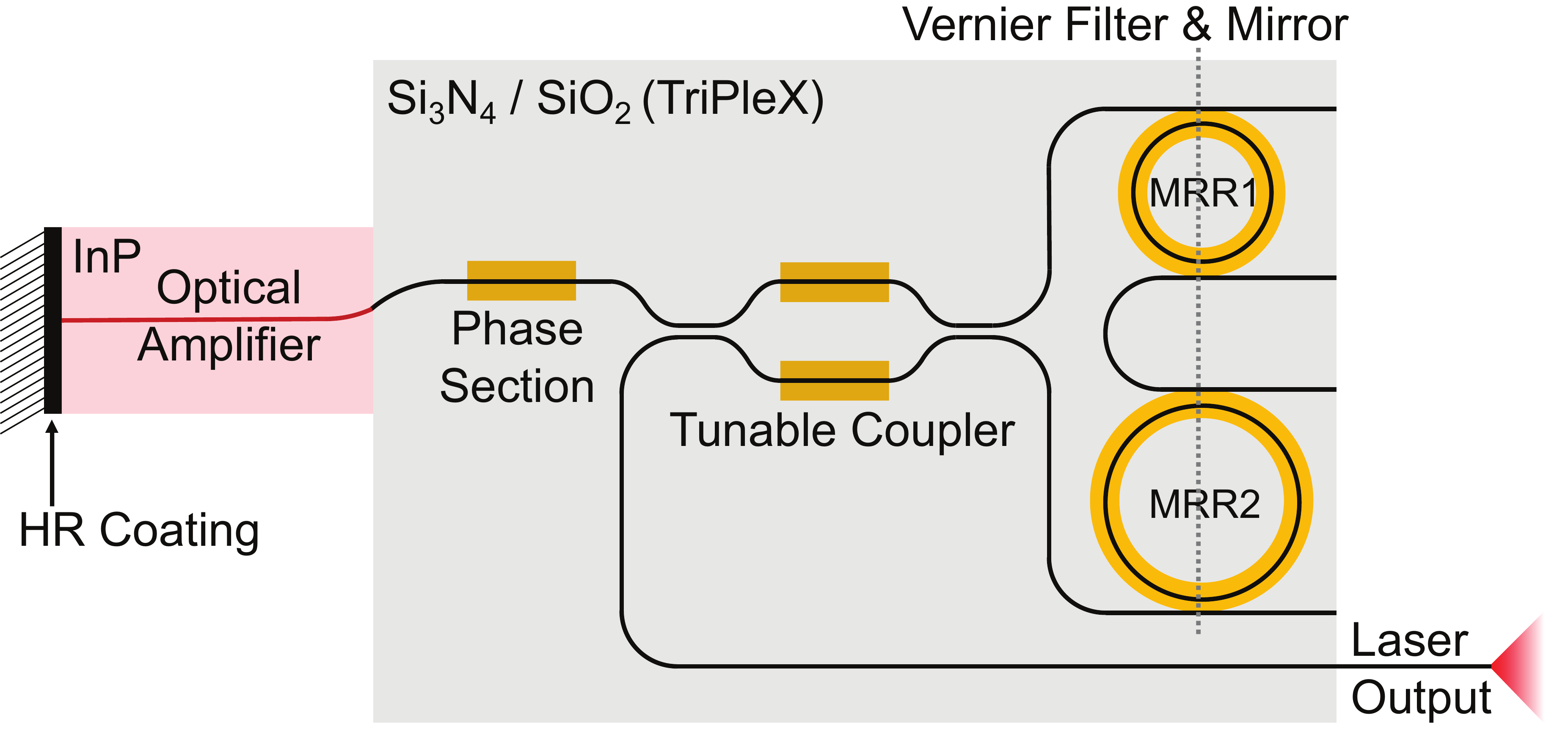}
    \includegraphics[width=0.33\textwidth]{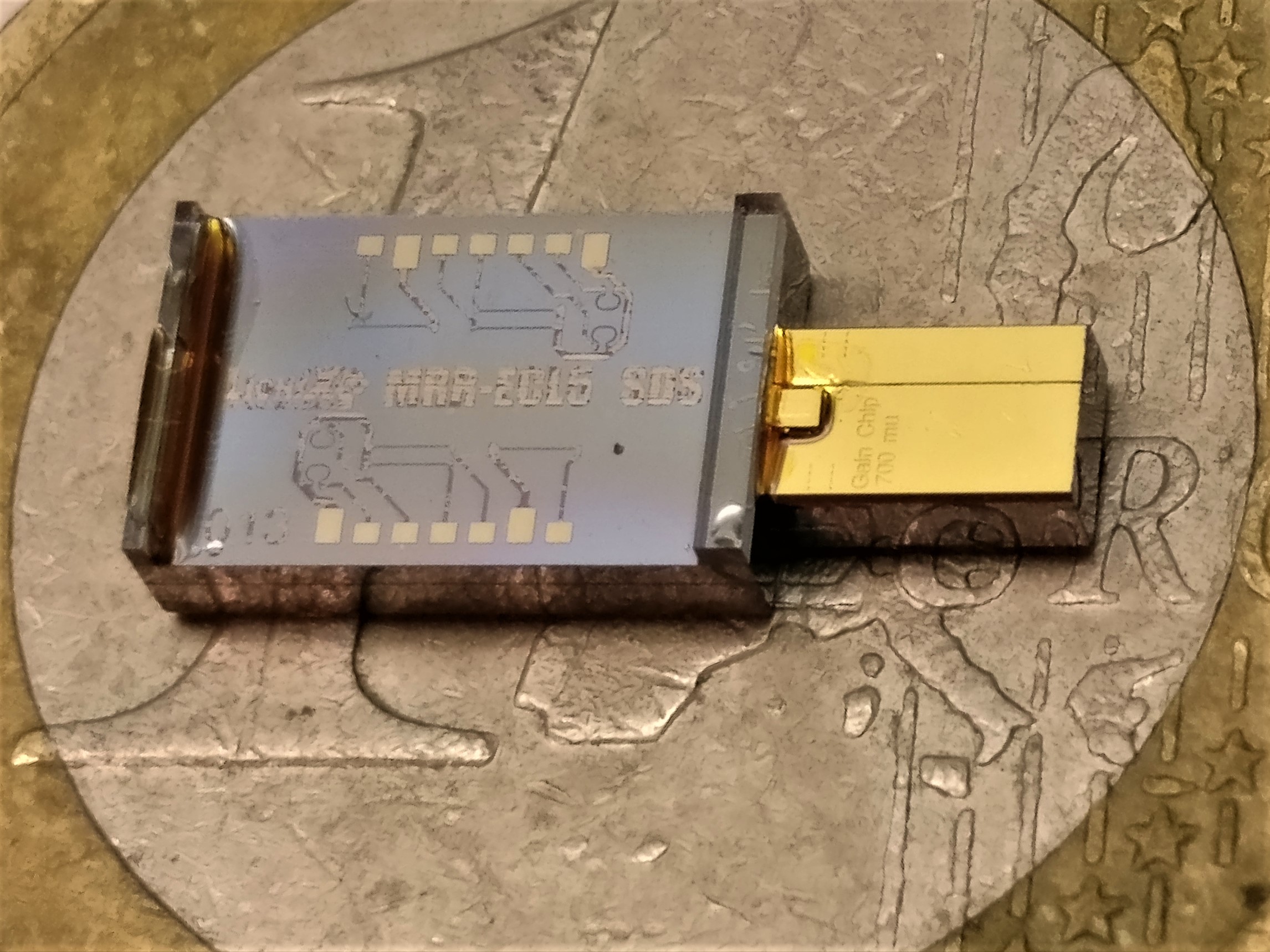}
    \caption{(a) Schematic view of the integrated laser. Shown are the semiconductor optical amplifier with a high reflectivity coating on one facet, the phase section, the two sequential microring resonators (MRR) in a loop mirror to provide wavelength selective feedback, and a tunable coupler to direct part of the light to the output port of the laser. The output of the laser is available via a fiber that is butt-coupled to the output port.
    (b)  Photo of the assembled hybrid laser, on top of a one Euro coin. The left part is the dielectric feedback chip with the heaters for thermal tuning, and the right part is the semiconductor amplifier chip on top of a submount.}
    \label{fig:laser_schematic}
\end{figure}

\section{Hybrid integrated extended cavity laser}

For an experimental demonstration, we use a hybrid integrated InP diode laser as shown in Fig.~\ref{fig:laser_schematic}~(a), with a waveguide circuit design similar to \cite{oldenbeuving_2013lpl, kobayashi_2015jlt, fan_2016pj, stern_2017ol, Komljenovic_2017as, lin_2018pj, mak_2019oe}. For single-frequency operation, wide wavelength tuning, and for narrow linewidth oscillation through cavity length extension, we use a low-loss Si$_3$N$_4$ waveguide circuit, which provides frequency-selective feedback. Figure~\ref{fig:laser_schematic}~(b) shows the hybrid assembly on top of a one Euro coin to indicate its small size.

The semiconductor chip~\cite{defilipe_2014ptl} is fabricated by the Fraunhofer Heinrich Hertz Institute and contains a multi-quantum well active waveguide based on InP with a single-pass geometric length of 700~$\mu$m and a gain bandwidth of at least 120~nm around 1540~nm. One facet has a high reflectance coating of $\sim$~90\% against air, which forms one of the mirrors of the laser cavity. At the other facet, light is coupled to the dielectric chip, which forms the other feedback mirror. To reduce undesired reflections at the interface between the two chips, an anti-reflection coating is applied to the facet and the waveguide is tilted by {9\textdegree} with respect to the facet normal.

The function of the dielectric chip is to provide frequency selective feedback and to increase the effective cavity length with very low loss. The feedback chip is based on a symmetric double-stripe waveguide geometry of two Si$_{3}$N$_{4}$ stripes buried in a SiO$_{2}$ cladding as described by Roeloffzen \textit{et al}~\cite{roeloffzen_2018jstqe}. This single-mode waveguide exhibits a low propagation loss of about 0.1~dB/cm and enables small bending radii down to 100~$\mu$m. We fully exploit the two-dimensional tapering capability of the Si$_{3}$N$_{4}$ platform for optimal matching of the optical mode to the mode of the gain chip and to the mode of the output optical fiber. At the facet for coupling with the InP gain chip, the waveguide is tapered two-dimensionally and angled down with {19.85\textdegree} with respect to the facet normal to match the InP waveguide optical mode and angle at the interface. This mode and angle matching allows efficient coupling and reduces spurious reflections to a minimum. 

A frequency selective filter is implemented on the feedback chip using two sequential race-track-shaped microring resonators, MRR$_{1}$ and MRR$_{2}$, in a Vernier configuration and placed inside a loop mirror (see Fig.~\ref{fig:laser_schematic}~(a)). The purpose of this filter is to impose single mode operation, tuning over the gain bandwidth and to increase the effective cavity length, which narrows the intrinsic linewidth~\cite{liu_2001apl}. The microrings have a circumference of 885.1~$\mu$m and 857.4~$\mu$m, respectively. These are the smallest lengths possible, based on the chosen implementation of racetrack resonators with adiabatic bends, the requirement of bend radii of at least 100~$\mu$m and the filter specifications. The total free-spectral range of the Vernier filter is 50.5~nm around the nominal wavelength. Although the bandwidth of the InP gain chip is larger, this free-spectral range is sufficient to obtain single mode lasing over a large part of the gain bandwidth. Both rings are symmetric and are designed for a power coupling coefficient to the bus waveguides of $\kappa^2=0.1$. We determined this value experimentally as $\kappa^2=0.071\pm0.003$, resulting in a length enhancement by a factor of 13.7 at resonance (\textit{cf.} Eq.~\ref{eq:Leff_resonant}). The length of the connecting bus waveguides and other elements adds up to 6.7~mm. In total this results in an effective optical cavity length of 3.5~cm.

Further, a so-called phase section of the bus waveguide is added to control the phase of the circulating light and to compensate phase changes when tuning any other element. Typically, the phase section is set to provide maximum feedback for a single longitudinal mode of the laser cavity, which restricts laser oscillation to a single wavelength and provides maximum output power. However, the phase section can also be tuned to provide equal feedback for two wavelengths and with this setting the laser can generate a multi-frequency comb~\cite{mak_2019oe}. 

A tunable coupler, implemented as a balanced Mach-Zehnder interferometer, is used to couple the circulating light out of the laser cavity. The extracted light is then directed to a single-mode polarization-maintaining output fiber. To prevent undesired external back reflections, the output fiber is terminated with an FC/APC connector and connected to a fiber isolator (Thorlabs IO-G-1550-APC).

Thermal tuning is implemented via resistive heaters placed above the rings, the phase section and the output coupler. The length of the phase section and output coupler heaters is 1~mm and both heaters require an electrical power of 290~mW to achieve a $\pi$ phase shift and can induce a change in the optical phase of at least $2.5\pi$. The slightly shorter heaters on top of the ring resonators require about 380~mW for $\pi$ phase shift and were tuned up to $1.6\pi$ phase shift during the experiments.

The amplifier, feedback chip and output fiber were all aligned for optimum coupling and fixed permanently. This hybrid assembly was mounted on a thermoelectric cooler in a 14-pin butterfly package. The cooler, amplifier and heaters are wire-bonded to the pins and connected to external drivers. This assembly of the hybrid laser enables stable laser operation, which is a prerequisite for accurate and reproducible wavelength tuning.

During the measurements presented here, the laser was operated with the following parameters, unless otherwise specified. The temperature of the thermoelectric cooler was set at 25~\textdegree C. Although this temperature is slightly above the temperature for maximum performance, it keeps the diode just above ambient temperature to avoid condensation and it reduces the optical output power only with a few percent. Furthermore, the output coupler was set to 80\% power outcoupling as this provides the best operating point for single-mode operation with high output power. Finally, after changing a laser parameter, the phase induced by the phase section was optimized for maximum output power, \textit{e.g.}, to compensate for changes in the roundtrip phase when the pump current is changed~\cite{buus_2005}. 

\section{Results}
\subsection{Basic laser properties}
\begin{figure}[bt]
    \centering
        \includegraphics[width=0.45\textwidth]{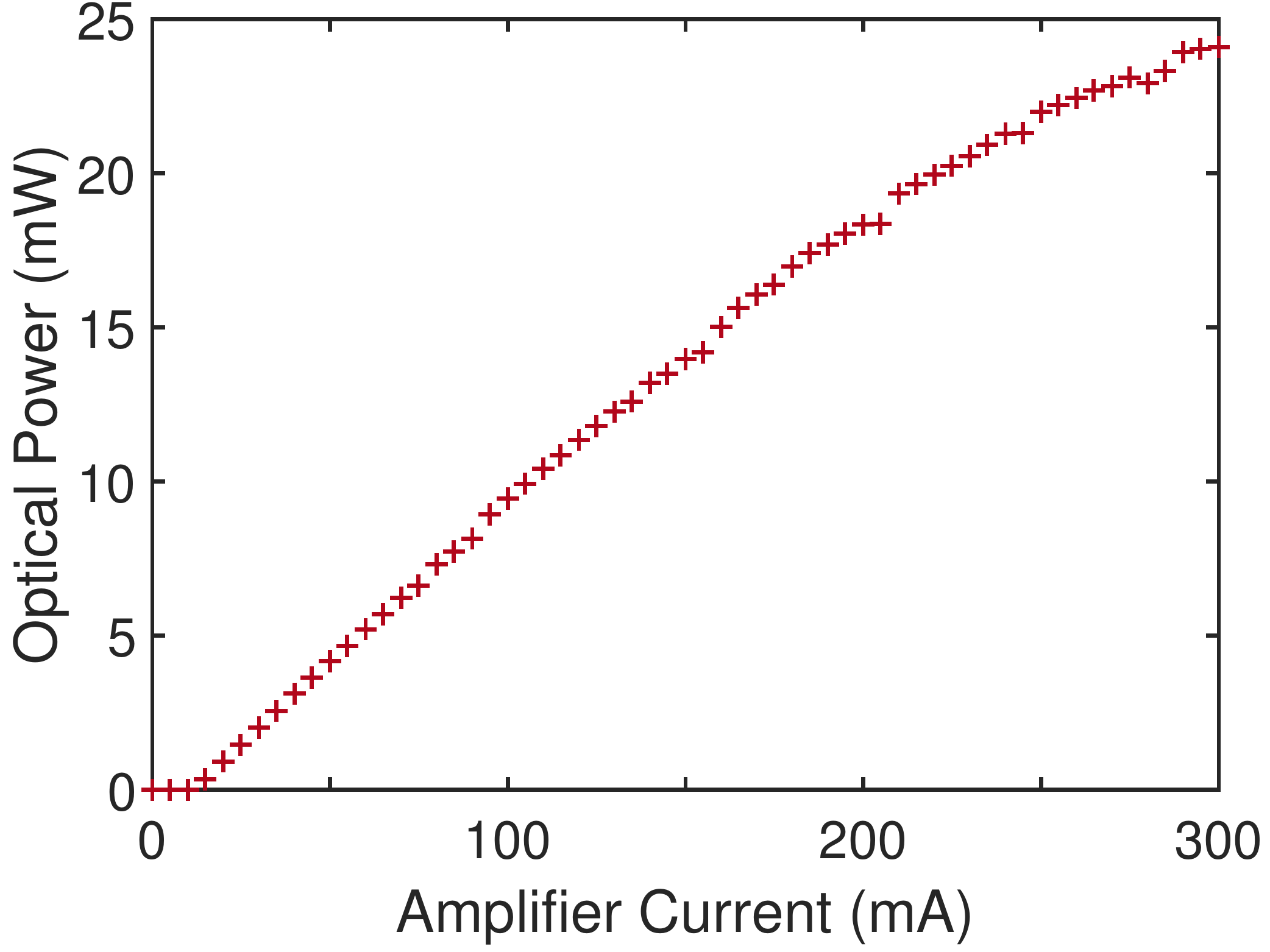}
    \caption{Measured fiber-coupled output power of the laser as function of the amplifier current. The thermoelectric cooler was set at 20~\textdegree C and the Vernier filter to a wavelength of 1576~nm.}
    \label{fig:curve_p_i}
\end{figure}

The fiber-coupled output power as a function of the pump current is shown in Fig.~\ref{fig:curve_p_i} with the Vernier filter set to a wavelength of 1576~nm and the temperature of the thermoelectric cooler to 20~\textdegree C, which are near optimum settings for this laser. A maximum of 24~mW was obtained at a pump current of 300~mA and the threshold current was 14~mA. 

\begin{figure}[bt]
    \centering
        \includegraphics[width=0.45\textwidth]{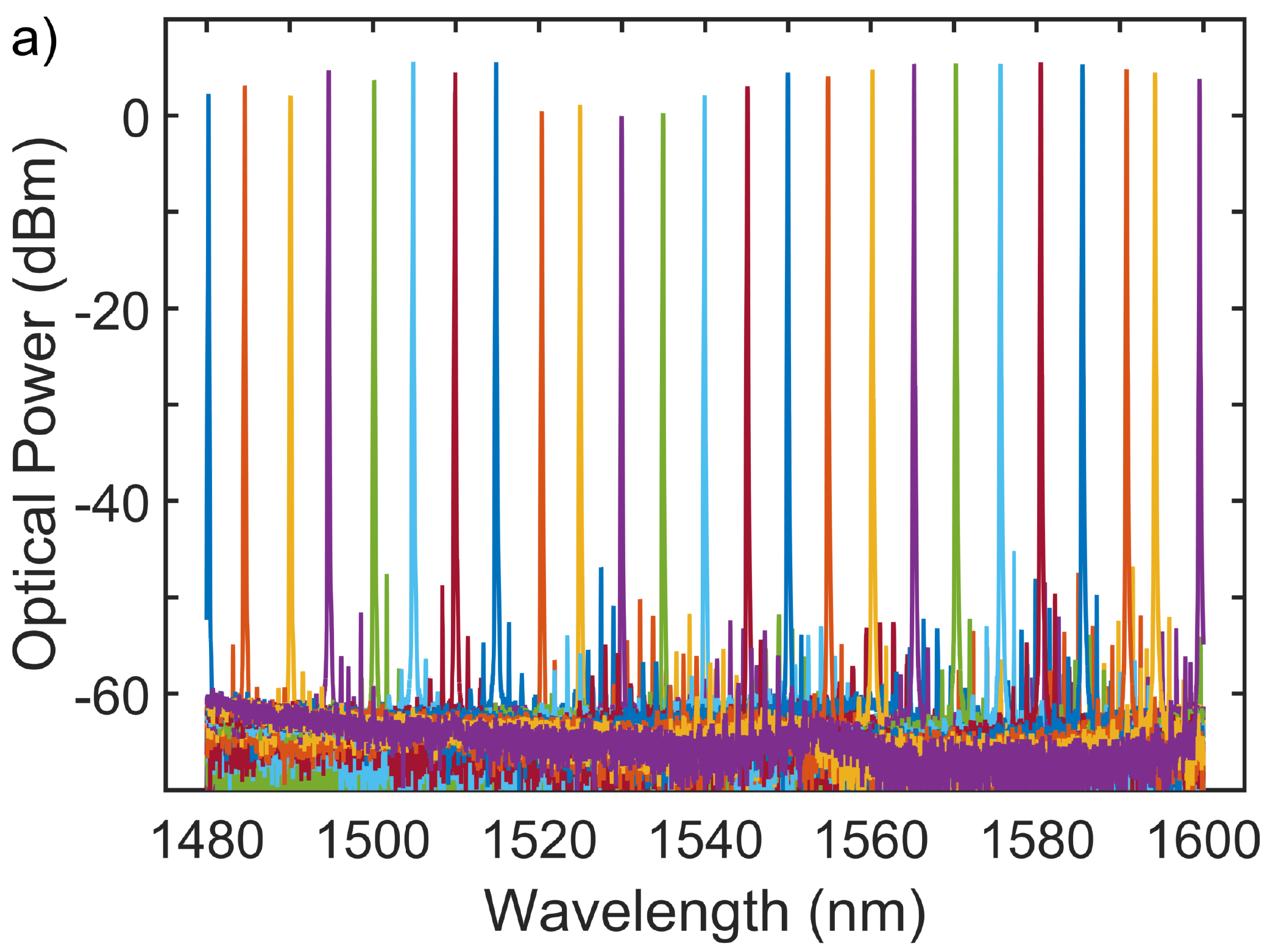}
        \includegraphics[width=0.45\textwidth]{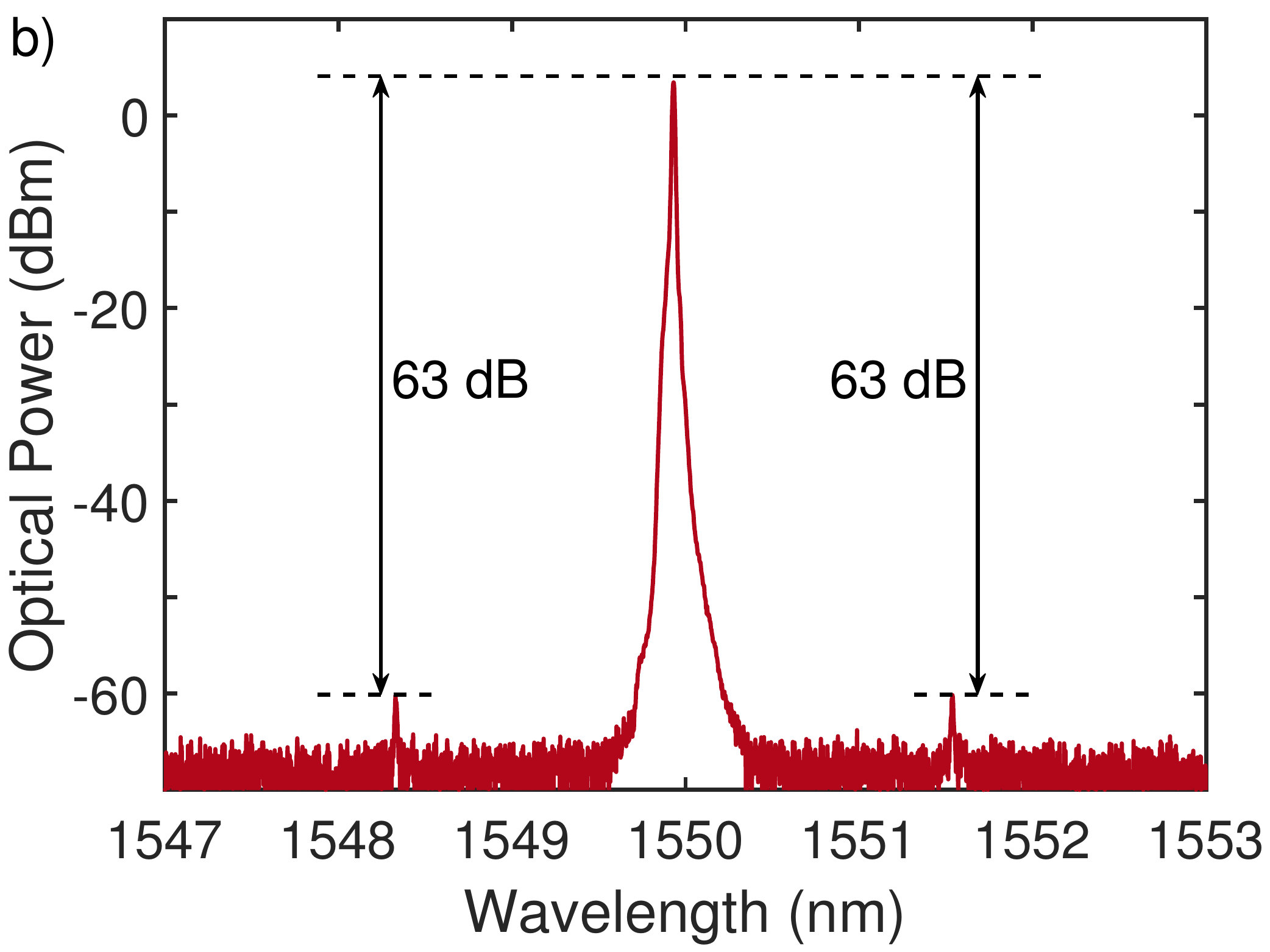}
    \caption{
        Superimposed laser spectra (a) showing the optical power in a 0.1~nm resolution bandwidth when the Vernier filter is tuned in steps of 5~nm, and measured laser power in a 0.01~nm resolution bandwidth (b), showing a high SMSR of 63~dB. The spectrum shown in (b) is an average over 10 measurements to reduce the background noise level and increase the visibility of the side modes. The amplifier current is set to 300~mA.}
    \label{fig:curve_tuning_smsr}
\end{figure}
To illustrate the broad spectral coverage of the laser, we show in Fig.~\ref{fig:curve_tuning_smsr}~(a) superimposed laser spectra, as measured with an ANDO AQ6317 optical spectrum analyzer (OSA1), when the Vernier filter is tuned in steps of approximately 5~nm. The pump current was increased to its maximum of 300~mA in order to obtain the broadest spectral coverage. Figure~\ref{fig:curve_tuning_smsr}~(a) shows a spectral coverage of 120~nm, \textit{i.e.}, extending the full gain bandwidth of the laser. The spectral coverage larger than the Vernier free-spectral range ($\sim$50~nm) was obtained by making use of the spectral dependence of the output coupler that can be varied. Optimum side mode suppression was found for a wavelength of 1550~nm. Figure~\ref{fig:curve_tuning_smsr}~(b) shows the optical power in a resolution bandwidth of 0.01~nm as a function of wavelength around 1550~nm, averaged over 10 measurements to bring the side modes above the noise level, again measured with OSA1. Figure~\ref{fig:curve_tuning_smsr}~(b) reveals two side modes with 63~dB suppression that are 1.61~nm away from the main mode, which agrees well with the first side peak of the Vernier filter. We used the high-resolution Finisar WaveAnalyzer 1500S optical spectrum analyzer (OSA2) to verify that only a single cavity mode was present.
We expected to find the highest side-mode suppression for wavelengths around 1570~nm, where the output power is maximum. However, with the Vernier filter tuned to that wavelength, another mode at one Vernier free-spectral range away ($\sim$50~nm), builds up as well, lowering the side-mode suppression. When the laser is tuned to 1550~nm, modes at $\sim$50~nm distance are not detectable with OSA1, possibly because these modes are suppressed by lower gain or higher losses.

\subsection{Intrinsic linewidth and mode-hop-free tuning}
\begin{figure}[bt]
    \centering
        \includegraphics[width=0.45\textwidth]{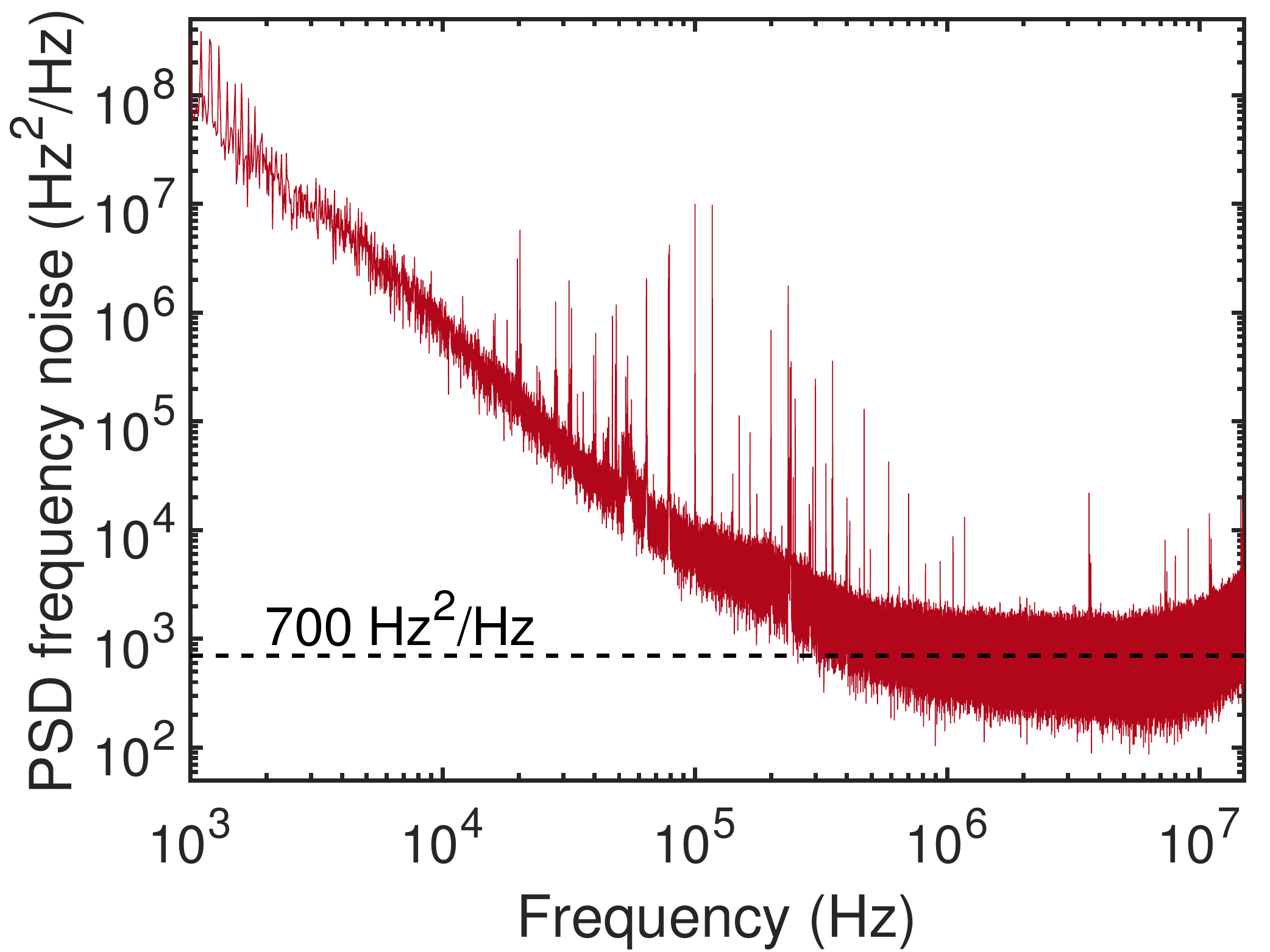}
    \caption{Measured single-sided power spectral density of the laser frequency noise. The dashed line at 700~Hz$^2$/Hz indicates the average noise for the noise frequency range between 1.3 and 3.5 MHz, excluding any spurious noise frequencies.}
    \label{fig:curve_psd}
\end{figure}

To demonstrate that the effective optical cavity length of 3.5~cm of the hybrid laser results in a high phase stability, we determined the intrinsic linewidth by measuring the power spectral density (PSD) of the frequency noise with a linewidth analyzer (HighFinesse LWA-1k 1550). This device was connected to the laser via the 10\% port of a 90:10 fiber optic coupler. The remaining 90\% was distributed over OSA1 and a photodiode. To obtain the lowest white noise level, we applied the maximum pump current of 300~mA from a battery-powered current source (ILX Lightwave LDX-3620) to the gain section. The laser wavelength was set to 1550~nm by the Vernier filter. Figure~\ref{fig:curve_psd} shows the measured PSD of the laser frequency noise, as function of the noise frequency. The frequency noise shown in Fig.~\ref{fig:curve_psd} has the characteristics of $1/f$-noise for noise frequencies below approximately 1~MHz and becomes white noise at higher frequencies. The narrow peaks in the spectral power density that can be observed at specific noise frequencies originate likely from either electronic
sources or RF-pickup in cables. The intrinsic linewidth is determined from the average white noise level, excluding spurious peaks, for noise frequencies between 1.3 and 3.5~MHz. Multiplying the single-sided PSD noise level of $700\pm230$~Hz$^2$/Hz with $\pi$ results in an intrinsic linewidth of $2.2\pm0.7$~kHz. This value is 4.5 times smaller than the previously reported value of 10~kHz for a similar InP-Si$_3$N$_4$ hybrid laser based on Vernier filters comprising two microring resonators~\cite{lin_2018pj}. We attribute this improvement mainly to a 4.6 times higher factor of pump current above threshold current in our measurement.

To verify continuous tuning and the increased mod-hop-free tuning sensitivity of the hybrid laser (see Eq.~\ref{eq:sensitivity}), we measured the laser wavelength and output power as a function of the electrical power applied to heater of the phase section with appropriate heater powers applied to the microring resonators (see Eq.~\ref{eq:syn_tun}). For these measurements, the laser was connected via a fiber optic coupler to OSA2 in order to resolve the small stepsize in wavelength and to a photodiode (Thorlabs~S144C) to monitor the output power. The amplifier current was set to 70~mA. For zero applied heater power to the phase section, we optimized the ring heaters to align the Vernier filter transmission with the lasing wavelength using minimum heater power. This resulted in an initial wavelength of 1534.25~nm and ensured having the full range of the phase section available for continuous tuning.

\begin{figure}[bt]
    \centering
        \includegraphics[width=0.45\textwidth]{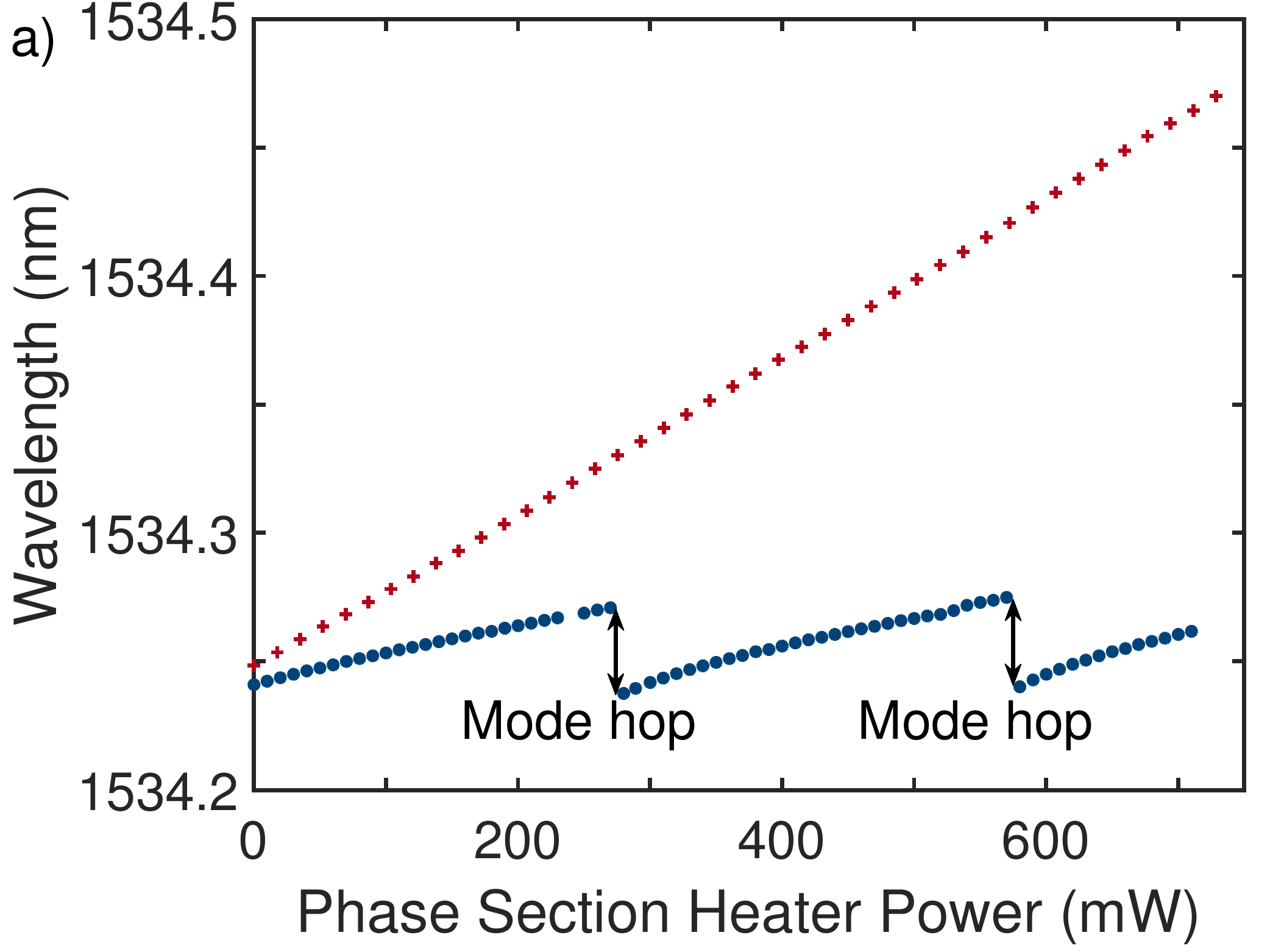}
        \includegraphics[width=0.45\textwidth]{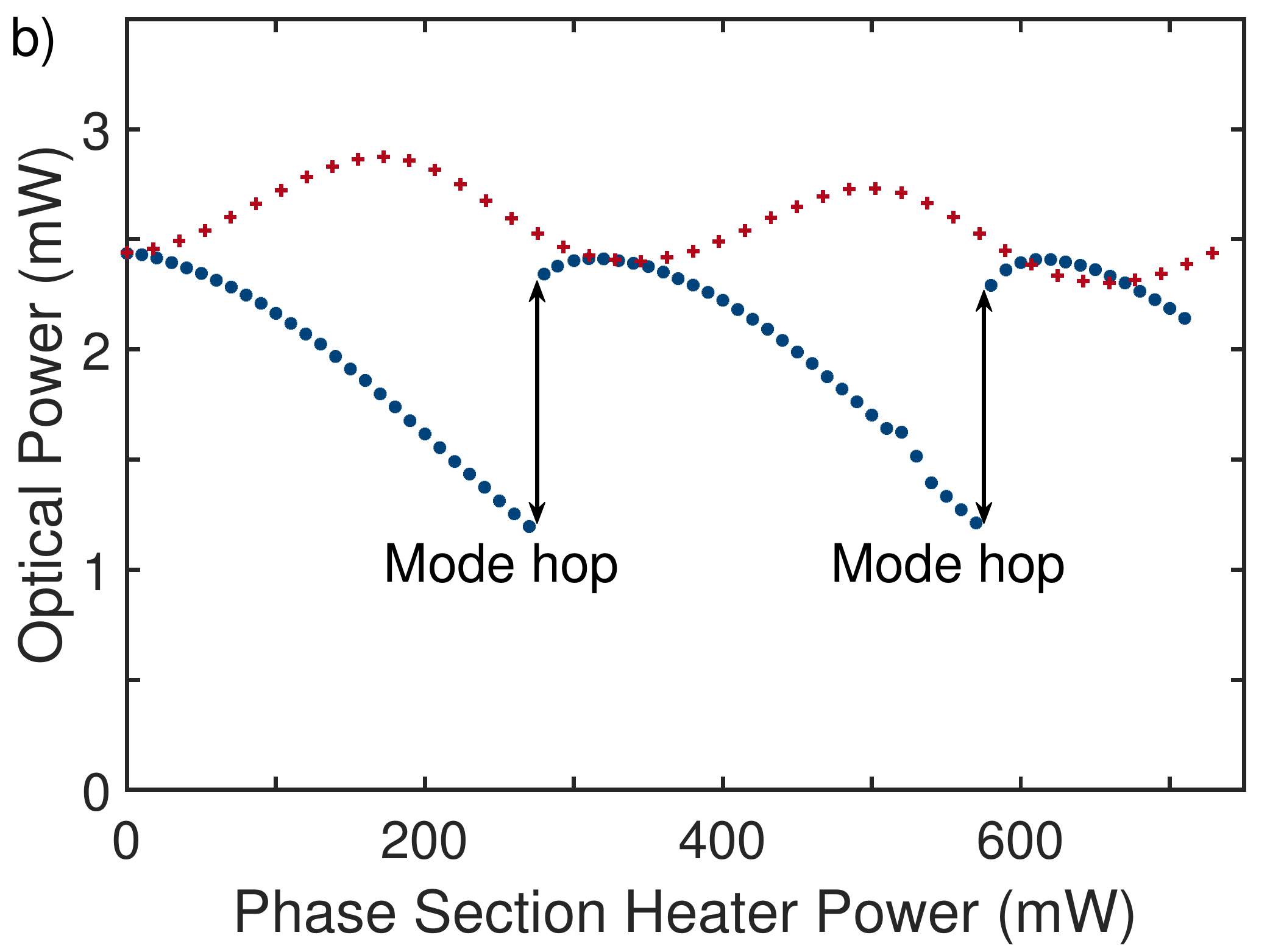}
    \caption{
        Measured laser wavelength (a) and output power (b) as function of the phase section heater power. Synchronous tuning of the phase section with the Vernier filter is shown by the red crosses, while tuning only the phase section is shown by the blue circles for comparison. Mode hops are indicated with arrows.}
    \label{fig:curve_tuning_mhf}
\end{figure}

For continuous tuning, we experimentally determined the optimum tuning ratios for the phase section versus the ring resonators by maximizing the laser output power as $\partial \phi_{1}/\partial\phi_{\textrm{ps}}=0.107$ and $\partial \phi_{2}/\partial\phi_{\textrm{ps}}=0.103$ for ring resonators 1 and 2, respectively. This agrees well with the values given by Eq.~\ref{eq:syn_tun} of 0.108 and 0.105, respectively, especially when considering the uncertainty in some of the laser parameters. Using these ratios, the heater power for the phase section was increased in steps corresponding to a change in lasing wavelength of 5~pm. The corresponding wavelength and output power are shown as red crosses in Figs.~\ref{fig:curve_tuning_mhf}~(a) and \ref{fig:curve_tuning_mhf}~(b), respectively. For comparison, Fig.~\ref{fig:curve_tuning_mhf} also shows the same measurements when only the heater of the phase section is varied while the heaters for the microrings are kept constant (blue circles). 

Figure~\ref{fig:curve_tuning_mhf}~(a) clearly shows that, when only the phase section is tuned, modes hops occur that limit the continuous tuning range to 0.034~nm, which is the free-spectral range of the laser cavity. In this case, the optical cavity length used to calculate the free-spectral range needs to include the effective optical lengths of the microring resonators. Furthermore, the optical power varies strongly and discontinuously during the tuning (see Fig.~\ref{fig:curve_tuning_mhf}~(b)). The reason is that even a small detuning between the lasing wavelength and the fixed position of the sharp Vernier filter transmission peak leads to a  strong change in the cavity losses and thus to a change in output power. Further tuning of the phase section, beyond integer multiples of $\pi$ phase shift, leads to mode hops, which can be observed as discontinuities in the wavelength and output power of the laser.

On the other hand, Fig.~\ref{fig:curve_tuning_mhf} shows no mode hops when the resonances of the microring resonators are synchronously tuned with the lasing wavelength via the phase section (red crosses). Although we aimed at constant optical power, there is still some remaining oscillation visible in Fig.~\ref{fig:curve_tuning_mhf}~(b) that corresponds in period with the distance between mode hops. We attribute this to a small changing mismatch between the lasing wavelength and the resonances of the ring resonators during the tuning. 

The most apparent difference in Fig.~\ref{fig:curve_tuning_mhf}~(a) is that the tuning sensitivity $\partial\lambda_{c}/\partial\phi_{\textrm{ps}}$ is indeed much larger for synchronous tuning compared to only tuning the phase section. We find that  $\partial\lambda_{c}/\partial\phi_{\textrm{ps}}=0.31$~pm/mW for synchronous tuning, while it is 0.11~pm/mW when only tuning the phase section. The tuning sensitivity of 0.31~pm/mW agrees well with the value of 0.29~pm/mW predicted by Eq.~\ref{eq:sensitivity}, given the uncertainty in some of the experimentally determined laser parameters. In addition, the synchronous tuning allows for a larger than $\pi$ phase shift induced by the phase section without invoking a mode hop. These effects together result in a total mode-hop-free tuning range of 0.22~nm, which is a six-fold increase over the free-spectral range of the laser cavity and only limited limit by the available phase shift of the phase section. 

\subsection{Acetylene absorption spectroscopy with mode-hop-free tuning}

In demonstrating the continuous tuning of the hybrid laser, we have applied wavelength increments of 5~pm that can easily be resolved using OSA2. However, it should be possible to tune the lasing wavelength in much smaller steps. Current resolution of the electronics used to power the heaters allows mode-hop-free laser wavelength tuning in steps below 0.1~pm, well beyond that what can be resolved by the available optical spectrum analyzers. We demonstrate this small step size by recording the shape of an acetylene absorption line in high resolution. Acetylene ($^{12}$C$_{2}$H$_{2}$) is chosen because it has several well-known sharp absorption lines in the wavelength range of interest, which can be modelled with high accuracy~\cite{goldenstein_2017jqsrt,rothman_2013jqsrt}. For this experiment, we choose the P19 ro-vibrational absorption line of acetylene with a center wavelength of 1536.713~nm. 

To measure the acetylene absorption line shape, we divided the laser output equally by a fiber optic coupler over two outputs. One part was sent through an absorption cell to a first photodiode (PD1, Thorlabs S144C), while the other part was split again and divided over a second photodiode of the same model (PD2) and OSA1 for calibration of the laser wavelength. The absorption cell is a standard sealed fiber-coupled acetylene gas cell from Wavelength Reference Inc., with a 5.5~cm path length and 50±5 Torr pressure at 295~K temperature. 

The normalized transmission through the absorption cell was determined by dividing the signal of PD1 by the signal of PD2 to compensate for any changes in output power of the laser and then normalize it to the maximum transmission measured in the total tuning range of \textasciitilde0.23~nm. Using the total tuning range, OSA1 was used to calibrate the increment in wavelength of the laser as 0.12~pm per 0.37~mW heater power applied to the phase section. However, The resolution of the OSA was insufficient to determine an accurate starting wavelength. This starting wavelength was determined by having the center wavelength of the measured P19 absorption line coincide with the value obtained from the calculated transmission. The required offset of -1.8~pm applied to the wavelength of the laser falls well within the resolution of OSA1. 

\begin{figure}[tb]
    \centering
    \includegraphics[width=0.45\textwidth]{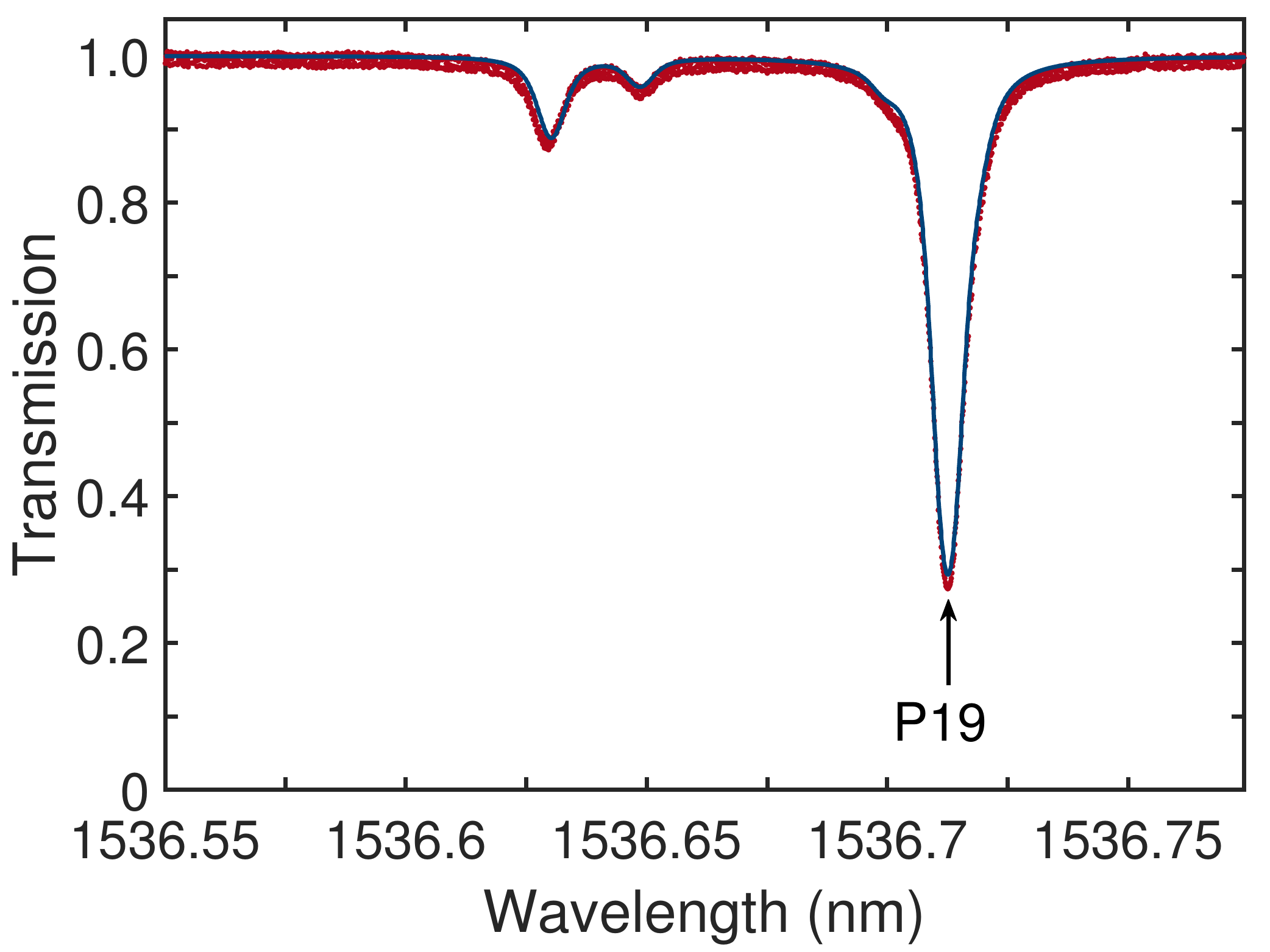}
    \caption{Measured (red dots) and calculated (blue line) transmission through an acetylene gas cell as function of wavelength for absorption line P19 at 1536.713~nm.}
    \label{fig:curve_acetylene}
\end{figure}
Figure~\ref{fig:curve_acetylene} shows the measured normalized transmission through the gas cell as function of the calibrated wavelength of the laser (red dots). The calculated normalized transmission~\cite{goldenstein_2017jqsrt} for nominal pressure corresponding to the temperature of the absorption cell is shown as a blue line.
We observe in Fig.~\ref{fig:curve_acetylene} an overall good match between the measured and calculated transmission. The slightly broader measured transmission can be explained by the uncertainty in gas cell pressure. By recording the P19 acetylene absorption line and comparing it with an accurately calculated spectrum, we show that, once calibrated, the hybrid extended cavity laser can be accurately tuned to any wavelength within the continuous tuning range of \textasciitilde0.23~nm (\textasciitilde29~GHz) with a resolution as small as \textasciitilde0.12~pm (\textasciitilde15~MHz), which is well below the resolution of both optical spectrum analyzers used.

\section{Discussion and conclusions}
We have presented a novel method for increasing the continuous tuning range of integrated single-frequency lasers with an ultra-narrow linewidth by increasing the tuning sensitivity $F_{\lambda}$. Ultra-narrow linewidth is provided by extending the cavity length with a  multi-pass resonator-based filter, as schematically shown as laser 3 in Fig~\ref{fig:cavities}~(a). In our demonstration experiment, the filter consists of two tunable microring resonators that not only enable single-mode oscillation, but also contribute to the optical length of the laser with multiple passes of the light through the rings per roundtrip in the laser cavity. When only tuning the phase section, the laser behaves like laser 2 in Fig~\ref{fig:cavities}~(a) and a small tuning sensitivity is found, corresponding to the long external laser cavity. However, when the intracavity filter is synchronously tuned with the phase section, the tuning sensitivity is significantly increased, corresponding to that of an equivalent laser with a short cavity without the effective length of the filter, \textit{i.e.} to laser 1 in Fig~\ref{fig:cavities}~(a).

We experimentally demonstrated the extended continuous tuning with a hybrid integrated semiconductor laser. The laser comprises a 700~$\mu$m long InP gain section, while the effective optical cavity length is extended to 3.5~cm, using a low-loss Si$_3$N$_4$ waveguide circuit, which narrows the intrinsic linewidth to 2.2~kHz. The circuit comprises a phase section for continuous tuning and a highly frequency selective Vernier filter with two tunable microring resonators. When only tuning the phase section, the tuning range per phase shift, $\partial\lambda_{c}/\partial\phi_{\textrm{ps}}$ is only 10~pm/rad. If the Vernier filter frequency is synchronously tuned with the oscillation frequency of the laser, the tuning sensitivity is enhanced by a factor of 2.8, to 28~pm/rad. 

The observed enhancement in tuning sensitivity agrees very well with the continuous tuning model we developed. This model shows that the enhanced tuning sensitivity relies on the sum of the optical length of the amplifier and the optical length of the bus waveguide. Since the tuning sensitivity is independent of the length of the microring resonators, this allows independent optimization of intrinsic phase stability and continuous tuning sensitivity. For further linewidth narrowing, the cavity length of a laser can be increased without a penalty in the range of continuous tuning, \textit{e.g.}, by increasing the ring diameter, lowering the bus to ring waveguide coupling constants or by adding extra rings~\cite{fan_2019arxiv,tran_2020jstqe}. The tuning sensitivity and hence the range of continuous tuning can be increased by shortening the length of bus waveguides not used as phase section. Alternatively, increasing the relative length of the bus waveguides used as phase section extends the continuous tuning range as well. For instance, the continuous tuning range of the laser investigated is limited by the maximum phase delivered by the phase section. If the heaters were extended to cover all of the available bus waveguides, this would increase the continuous tuning range by a factor of 6.7 to approximately 1.6~nm, while maintaining a 2.2-kHz~linewidth.

Phase shifters that are more effective than the current $2.5\pi$~phase shift per mm length would allow for laser designs with continuous tuning over a much larger range, \textit{i.e,} beyond 2~nm. 

An additional advantage of the synchronous phase-and-filter tuning is that it can mitigate technological limitations in laser wavelength tuning, wavelength modulation, or frequency locking to references. The reason is that the available phase shift of a tunable phase section in integrated photonic circuits is often limited by material constants, \textit{e.g.}, the thermo-optic, strain-optic, or electro-optic coefficients of the involved materials. Our approach increases the tuning sensitivity $F_{\lambda}$, which means that less phase shift becomes sufficient for a given tuning range. This is an advantage in many respects, such as to reduce power consumption in thermal phase shifters and to increase the speed in laser tuning and frequency modulation, for facilitating, \textit{e.g.}, spectroscopic and metrologic applications.

The option of faster modulation and frequency control is specifically important for stabilizing lasers to reference frequencies, for instance to absolute standards~\cite{deLabachelerie_1995ol}. We experimentally demonstrated fine-tuning of the laser in steps of 0.12~pm over a full P19 ro-vibrational absorption line of acetylene. Because the recording revealed the known line shape~\cite{rothman_2013jqsrt,goldenstein_2017jqsrt} very accurately, we conclude that the laser can be continuously and linearly tuned with steps smaller than the resolution of the available optical spectrum analyzers. Furthermore, such tuning is essential for stabilization of hybrid integrated InP-Si$_3$N$_4$ lasers to an absolute reference. 

We expect that the demonstrated tuning method based on resonant intracavity filtering can be applied to other types of lasers as well. Possible examples are photonic crystal lasers~\cite{tandaechanurat_2011NP,takahashi_2013N}, Bragg waveguides~\cite{numata_2010OE,huang_2019o}, other gain materials~\cite{grivas_2016PQE,kiani_2019OL} or nonlinear gain. The expected advantages are linewidth narrowing without obstructing continuous tuning, increasing the tuning range without obstructing linewidth narrowing, and reduced power consumption or higher speed in continuous tuning or stabilization of single-frequency lasers.

\section*{Appendix - Mode-hop-free tuning}
\renewcommand{\theequation}{A-\arabic{equation}}  % redefine the command that creates the equation no.
\setcounter{equation}{0}  % reset counter 
The hybrid laser produces maximum output power when each of the ring resonators has a resonant wavelength $\lambda_{i}$ ($i=1,2)$ that coincides with the lasing wavelength $\lambda_{c}$. Mode-hop-free tuning requires that the shift in each of the wavelengths is the same, \textit{i.e.}, $\delta\lambda_{c}=\delta\lambda_{1}=\delta\lambda_{2}$. Tuning of the resonant wavelengths is realized via resistive heaters on top of the phase section and the microring resonators. Heating of the phase section changes only the lasing wavelength, while heating a microring resonator will concurrently change the resonant wavelength of the ring and the lasing wavelength. As we will demonstrate, it is the latter property that causes the mode-hop-free tuning sensitivity, $\frac{\partial\lambda_{c}}{\phi_{\textrm{ps}}}$, of this hybrid laser with resonantly enhanced optical cavity length to be larger than that of a standard semiconductor Fabry-P{\'e}rot laser of equivalent optical length. 

In order to model the action of the heaters, we assume that their sole action is to add a phase, $\phi_{j}$, to the light propagating through the waveguide section $j$ below the heating element. As the free-spectral range of a resonator is defined as an increase in roundtrip phase of the light with $2\pi$ it is intuitive and straightforward to show that the change in resonant wavelength,  $\lambda_{i}$, for microring resonator, $i$, with phase added by the heater on top of the ring, $\phi_{i}$, is given by
\begin{equation}
    \frac{\partial \lambda_{i}}{\partial \phi_{i}}=\frac{\Delta\lambda_{i}}{2\pi},
    \label{eq:dldp_rings}
\end{equation}
where $\Delta\lambda_{i}$ is the free-spectral range of ring $i$. This free-spectral range is given by~\cite{rabus_2007} 
\begin{equation}
    \Delta\lambda_{i}(\lambda) = \frac{\lambda^2}{n_{g,i}(\lambda)L_{i}},
    \label{eq:fsr_rings}
\end{equation}
where $\lambda$ is the wavelength in vacuum, $n_{g,i}$ is the effective group index of the Si$_{3}$N$_{4}$ ring waveguide and $L_{i}$ is the circumference of the ring resonator. In a similar way, the change of lasing wavelength, $\lambda_c$, with phase added by the phase section, $\phi_{\textrm{ps}}$, is given by
\begin{equation}
    \frac{\partial \lambda_{c}}{\partial \phi_{\textrm{ps}}}=\frac{\Delta\lambda_{c}}{\pi},
    \label{eq:dldp_cavity}
\end{equation}
as the light passes through the phase section twice per roundtrip. In Eq.~\ref{eq:dldp_cavity}, $\Delta\lambda_{c}$ is the free-spectral range of the laser cavity given by
\begin{equation}
    \Delta\lambda_{c}(\lambda) = \frac{\lambda^2}{2 n_{g,a}(\lambda)L_{a}+2 n_{g,b}(\lambda)L_{b}+\sum_{i}n_{g,i}(\lambda)L_{e,i}(\lambda)},
    \label{eq:fsr_cavity}
\end{equation}
where $n_{g,j}$ is the effective group velocity of the light in the semiconductor waveguide ($j=a$), the Si$_{3}$N$_{4}$ bus waveguide ($j=b$) or the Si$_{3}$N$_{4}$ rings ($j=i$), $L_{j}$ is the geometric length of section $a$ or $b$ and $L_{e,i}$ is the effective length of ring $i$ given by~\cite{liu_2001apl}
\begin{equation}
    L_{e,i}(\lambda) = -\frac{1}{2\pi}\frac{\lambda^{2}}{n_{g,i}(\lambda)}\frac{\partial \theta_{i} }{\partial \lambda},
    \label{eq:Leff}
\end{equation}
with $\theta_{i}$ the phase added to the light when it propagates through the ring from input to drop port. For a symmetric microring resonator at resonance, $L_{e,i}$ reduces to~\cite{liu_2001apl}
\begin{equation}
    L_{e,i} = L_{i}\left(\frac{1}{2}+\frac{1-\kappa^2}{\kappa^2} \right),
    \label{eq:Leff_resonant}
\end{equation}
where $\kappa^2$ is the power coupling from the input waveguide to the ring and from the ring to the output waveguide. The effective length is equal to the geometric length $L_{i}/2$ from input to drop port to which a length is added, equal to the distance travelled by the light in the time it effectively remains within the ring, which corresponds to the maximum attainable effective length of the ring.

The laser wavelength is not only shifted by the phase induced by the phase section (Eq.~\ref{eq:dldp_cavity}), but also by the phase induced by the heaters on top of the rings. At resonance, the total phase added to the light in the laser cavity by ring $i$ is just $\phi_{c,i}= \frac{L_{e,i}}{L_{i}}\phi_{i}$, which results in a change in lasing wavelength of
\begin{equation}
    \frac{\partial \lambda_{c,i}}{\partial \phi_{i}}=\frac{\Delta\lambda_{c}}{2\pi}\frac{L_{e,i}}{L_{i}}.
    \label{eq:dlcdp_rings}
\end{equation}
For the total change in lasing wavelength, $\delta\lambda_{c}$, we then have  
\begin{equation}
    \delta\lambda_{c}=\frac{\partial \lambda_{c}}{\partial \phi_{\textrm{ps}}}\delta\phi_{\textrm{ps}}+\sum_{i}\frac{\partial \lambda_{c,i}}{\partial \phi_{i}}\delta\phi_{i}.
    \label{eq:dlc}
\end{equation}
The mode-hop-free condition $\delta\lambda_{c}=\delta\lambda_{i}$ requires that the phase added to the light in the rings is proportional to the phase added by the phase section. Using this and substituting Eqs.~\ref{eq:dldp_rings}, \ref{eq:fsr_rings}, \ref{eq:dldp_cavity} and \ref{eq:dlcdp_rings} in Eq.~\ref{eq:dlc} gives 
\begin{equation}
    \delta\lambda_{c}=\frac{\Delta\lambda_{c}}{\pi}\delta\phi_{\textrm{ps}} + \sum_{i}n_{g,i}L_{e,i}\frac{\Delta\lambda_{c}}{\lambda_{c}^{2}}\delta\lambda_{c}.
    \label{eq:dlc_final}
\end{equation}
Using Eq.~\ref{eq:fsr_cavity} in Eq.~\ref{eq:dlc_final} gives the mode-hop-free change in lasing wavelength with phase induced by the phase section, which we have introduced as the tuning sensitivity $F_{\lambda}$ in the main text, as
\begin{equation}
    F_{\lambda}\equiv\frac{\partial\lambda_{c}}{\partial\phi_{\textrm{ps}}}=\frac{1}{\pi}\frac{\lambda_{c}^{2}}{2n_{g,a}L_{a}+2n_{g,b}L_{b}},
    \label{eq:tuning_sensitivity}
\end{equation}
Equation~\ref{eq:tuning_sensitivity} shows indeed that the mode-hop-free tuning sensitivity does not include the effective length of the microring resonators that are part of the laser cavity, and the tuning sensitivity corresponds to that of a laser with a short cavity length.

To satisfy the mode-hop-free tuning condition, the phase added by the ring resonator heaters must be set with a fixed ratio to the phase added by phase section heater. Using Eqs. \ref{eq:dldp_rings}, \ref{eq:fsr_rings} and \ref{eq:tuning_sensitivity} gives this ratio as
\begin{equation}
    \frac{\partial \phi_{i}}{\partial \phi_{\textrm{ps}}}=\frac{n_{g,i}L_{i}}{n_{g,a}L_{a}+n_{g,b}L_{b}}.
    \label{eq:syn_tun}
\end{equation}
As the right-hand-side of Eq.~\ref{eq:syn_tun} is typically less than 1 for most laser designs, the phase added by the ring resonator heaters is usually less than that by the phase section heater. Consequently, the total continuous tuning range is often limited by the phase section heater.

\section*{Funding}
This research is funded by the European Union's Horizon 2020 research and innovation programme under grant agreement 780502 (3PEAT)

\section*{Acknowledgments}
The authors would like to thank A.~van de Kraats for support with the driver electronics, M.~Hoekman for the mask design, R.~E.~M.~Lammerink for assistance with the linewidth measurements and J.~Mak for useful discussions.

\section*{Disclosures}
The authors declare no conflicts of interest.

\bibliographystyle{osajnl}

\end{document}